\documentclass[usenatbib]{mn2e}
\usepackage[latin1]{inputenc}
\usepackage{amsmath}
\usepackage{amsfonts}
\usepackage{amssymb}

\usepackage{graphicx}

\def\fnl {f_{\rm{NL}}}
\def\gnl {g_{\rm{NL}}}
\def\be {\begin{equation}}
\def\ee {\end{equation}}
\def\db {\Delta b(k,\fnl,\gnl)}

\def\Pm {P_{\mathrm{m}}}
\def\Pc {P_{\mathrm{hm}}}
\def\OM {\Omega_{\mathrm{m}}}
\def\iMPC {\, h\, \mathrm{Mpc}^{-1}}
\def\Msun {\, h^{-1}\, \mathrm{M}_{\odot}}
\def\MPC {\, h^{-1}\, \mathrm{Mpc}}
\def\nut {\tilde{\nu}}

\begin{document}

\title[Can we measure $\fnl$ from the power spectrum?]{Can we really measure $\fnl$ from the galaxy power spectrum?}
\author[N. Roth, C. Porciani]{Nina Roth\thanks{E-mail: nroth@astro.uni-bonn.de}, Cristiano Porciani
\newauthor
\footnotesize{Argelander-Institut f\"ur Astronomie der Universit\"{a}t Bonn, Auf dem H\"ugel 71, D-53121 Bonn, Germany}}

\maketitle
\begin{abstract}
The scale-dependent galaxy bias generated by primordial non-Gaussianity (PNG) can be used to detect and constrain deviations from standard single-field inflation. The strongest signal is expected in the local model for PNG, where the amplitude of non-Gaussianity can be expressed by a set of parameters ($\fnl$, $\gnl$, ...). Current observational constraints from galaxy clustering on $\fnl$ and $\gnl$ assume that the others PNG parameters are vanishing. Using two sets of cosmological $N$-body simulations where both $\fnl$ and $\gnl$ are non-zero, we show that this strong assumption generally leads to biased estimates 
and spurious redshift dependencies of the parameters. Additionally, if the signs of $\fnl$ and $\gnl$ are opposite, the amplitude of the scale-dependent bias is reduced, possibly leading to a false null detection. Finally we show that model selection techniques like the Bayesian evidence can (and should) be used to determine if more than one PNG parameter is required by the data.
\end{abstract}

\section{Introduction}
\label{sec:intro}
Standard single-field inflation scenarios predict primordial perturbations with nearly Gaussian initial conditions.
In fact, the tiny non-Gaussian features arising from the self-coupling of the inflaton field would remain undetectable. Any evidence for primordial non-Gaussianity (PNG) would then imply the emergence of non-standard physics in the early universe (for a review see e.g. \citealt{2004PhR...402..103B}).

Here we consider the local model of PNG, in which the Bardeen curvature potential during matter domination (at high-redshift) is given by
\begin{align}
\Phi(\mathbf{x}) = \phi(\mathbf{x}) & + \fnl \left[\phi(\mathbf{x})^2-\langle \phi^2 \rangle \right] \nonumber \\
 & + \gnl \left[\phi(\mathbf{x})^3-3\langle \phi^2 \rangle \phi(\mathbf{x}) \right] ,
\label{png} 
\end{align}
where $\phi(\mathbf{x})$ is a zero-mean Gaussian random field, $\fnl$ and $\gnl$ are real constants, and the symbol $\langle \cdot \rangle$ denotes expectation values. On large scales, PNG of the local type introduces a scale-dependent bias between galaxies and the underlying matter distribution (\citealt{2008PhRvD..77l3514D}, \citealt{2008ApJ...677L..77M}, \citealt{2008JCAP...08..031S}, \citealt{2010PhRvD..81f3530G}). Several authors have derived observational constraints on $\fnl$ and $\gnl$ from the galaxy power spectrum or 2-point correlation function (\citealt{2008JCAP...08..031S}, \citealt{PhysRevD.81.023006}), and recently \cite{2011JCAP...08..033X} reported a possible detection of positive $\fnl$ at the $2\sigma$ level. However, these studies all assume that the underlying model for PNG is either purely quadratic ($\gnl = 0$) or purely cubic ($\fnl = 0$). In this Letter, we show that the assumption of a one-parameter model significantly biases the estimation of the PNG parameters when both $\fnl$ and $\gnl$ do not vanish. We also illustrate that model-selection techniques like the Bayes factor are powerful tools to identify the model that best describes the available data. This is particularly interesting in light of the next generation of galaxy and cluster surveys like eROSITA \citep{2010SPIE.7732E..23P}, \textit{Euclid} \citep{2011arXiv1110.3193L} or LSST \citep{2008arXiv0805.2366I} which are expected to significantly reduce the error bars of PNG parameters (e.g. \citealt{2012MNRAS.422...44P}, \citealt{2012MNRAS.tmp.2888G}).
\section{Scale-dependent bias and PNG}
\label{sec:db}
The linear matter density contrast in Fourier space $\delta_{\rm{m}}(k)$ follows from the curvature perturbations by
\be 
\delta_{\rm{m}}(k,z)= \alpha(k,z) \Phi(k) ,
\label{dfromp}
\ee
with
\be 
\alpha(k,z)=\frac{2 k^2 D(z) T(k)}{3 \OM H_0^2} ,
\label{alpha}
\ee
where $\OM$ and $H_0$ are the current matter-density parameter and Hubble constant, $T(k)$ is the transfer function, and $D(z)$ is the 
growth factor normalised to $(1+z)^{-1}$ during the epoch of matter domination.

Non-linear structures form with time via gravitational instability. Let us denote by $\Pm(k)$ the matter power spectrum and by $\Pc(k)$
the cross spectrum between dark-matter haloes (of a given mass) and mass. On large scales, the relation between $\Pc(k)$ and $\Pm(k)$ is given by 
\be 
\frac{\Pc(k)}{\Pm(k)}= b_1(\fnl,\gnl)+ \db ,
\label{pmph}
\ee
where $b_1$ is the linear bias parameter (a real-valued constant) and $\db$ is a scale-dependent bias term introduced by PNG. Following the work of \cite*{2011arXiv1106.0503S} (hereafter SFL12), the latter term can be split up into different contributions (to first order in $\fnl$ and $\gnl$):
\be 
\db = \frac{\beta_f \fnl + \beta_g \gnl}{\alpha(k)} .
\label{dbtot}
\ee
Its dependence on $\fnl$ was shown to be (\citealt{2008PhRvD..77l3514D}, \citealt{2008ApJ...677L..77M}, \citealt{2008JCAP...08..031S}, \citealt{2010PhRvD..81f3530G}):
\be
\beta_f = 2\,(\nut^2-1) ,
\label{dbf} 
\ee
with
\be
\nut \equiv \left[ \delta_c (b_1-1) +1 \right]^{1/2} 
\label{dalal}
\ee
where $\delta_c$ is a parameter set to $\sim 1.4$ to fit the clustering properties of friends-of-friends halos in numerical simulations. 
Note that $b_1$ modulates the amplitude of the scale-dependent bias through $\nut$. \footnote{Some authors write $\nut$ in term of the linear-bias factor obtained from Gaussian initial conditions, but \cite{2010PhRvD..81f3530G} have demonstrated that the full $b_1$ should appear in equation (\ref{dalal}). Note, however, that the correction to $b_1$ due to PNG is generally small and, in most practical cases, the difference between the two formulations is negligible.}
Similarly, the dependence on $\gnl$ is (SFL12)
\begin{align}
\beta_g =&\ \kappa_3^{(1)}(M) \left[ -0.7 +1.4(\nut -1)^2 +0.6(\nut-1)^3 \right] \nonumber \\ &\ -\frac{d \kappa_3^{(1)}(M)}{d \log \sigma^{-1}} \left( \frac{\nut-\nut^{-1}}{2} \right)
\label{beta}
\end{align}
with $\kappa_3^{(1)}$ the skewness of the mass density field (smoothed on the mass scale $M$) in the case of $\fnl=1$ and $\sigma$ the linear rms 
mass-density fluctuation on the scale $M$.
The skewness and its derivative can be expressed in terms of $b_1$ and $z$ (see equations 46 and 47 in SFL12).

In brief, PNG of the local type makes the bias of dark-matter haloes scale dependent and proportional to $k^{-2}$ for $k\to 0$.
It is therefore conceivable to constrain PNG from measurements of the galaxy power spectrum, provided a model that links galaxies to haloes.
However, the $\fnl$ and $\gnl$ contributions to the scale-dependent bias in  equation (\ref{dbtot}) show the same scaling with the wavenumber and it is not
possible to measure both parameters from studies of a single population of tracers of the large-scale structure.
Given that $\beta_f$ and $\beta_g$ display different scaling with halo mass and redshift, this degeneracy can be broken by considering several populations
of galaxies, possibly within widely separated redshift bins. Recent studies have accomplished this task by combining a number of observational datasets with simple models where either $\fnl$ or $\gnl$ are set to zero
a priori. With the help of numerical simulations, in this Letter we study the effect of these simplifying assumptions on the determination of PNG.
\section{Simulations}
\label{sec:sim}
We have used \textsc{Gadget-2} \citep{2005MNRAS.364.1105S} to perform two sets of dark-matter-only simulations consisting of four realisations each. Each set corresponds
to a different fiducial model of PNG assuming $\fnl= 50$ and $\gnl= \pm \, 5 \cdot 10^5$, in accordance with upper limits obtained from cosmic-microwave-background studies (\citealt{2011ApJS..192...18K}, \citealt{2010MNRAS.404..895V}). 
The other cosmological parameters are chosen in concordance with the WMAP5 analysis \citep{2009ApJS..180..330K}: $\Omega_{\rm{m}}=0.279$, $\Omega_{\Lambda}=0.721$, $\Omega_{\rm{b}}=0.0462$, $\sigma_8=0.817$, $h=0.7$, $n_{s}=0.96$. 
We follow $1024^3$ particles within a cubic box of side $L=1200 \MPC$, corresponding to a particle mass of $1.247 \cdot 10^{11} \Msun$. 
The initial conditions are generated at redshift 50 using the Zel'dovich approximation to displace particles from a regular Cartesian grid.

Haloes are selected using a friends-of-friends halo finder with a linking length of 0.2 times the mean interparticle separation. 
We consider only objects with at least 80 particles, so the halo mass range is $10^{13} \Msun \leq M_{\rm{halo}} 
\leq 2 \cdot 10^{15} \Msun$ at $z=0$. We use outputs at $z=\{0, 0.5, 1, 1.5, 2\}$. 
The scale-dependent bias parameter $b(k)$ for the haloes is measured from the ratio of the ($k$-binned) halo-matter cross spectrum and the matter auto spectrum, 
averaged over the four realisations. We only consider the range $0.007 \iMPC \le k \le 0.04 \iMPC$ where equation (\ref{beta}) holds as shown in SFL12. The errors 
on $b(k)$ are calculated using an estimator derived in the appendix of \cite*{2011JCAP...11..009S}. 

To make sure that equations (\ref{dbtot})-(\ref{beta}) (with $\delta_c=1.42$) can also be applied to our simulations, 
we fix $\fnl$ and $\gnl$ to the input values and use a standard $\chi^2$-fit to measure $b_1$ as a function of halo mass and redshift. We find good $\chi^2$ values in all cases, and the fitted $b_1$ is compatible with the value expected for the case of Gaussian initial conditions (see equation 14 from \citealt{Pillepich08}). We can therefore safely assume that the fitting formulae given by SFL12 accurately describe the scale-dependent bias for our sets of simulations.
\begin{table}
\centering
\caption{The top section describes the galaxy samples used in the analysis of Xia et al. (2011): redshift, type, effective linear bias parameter $b_{\rm{eff}}$ and number of objects $N_{\rm{gal}}$. The middle section illustrates the properties of specific halo subsamples from our simulations that have been selected
to approximate the observational dataset described above: redshift, mass range (in $10^{13} \Msun$), linear bias parameter $b_1$, number of objects in each mass bin, $N_{\rm{halo}}$,
after combining the four realisations. We denote this as the ``current sample''.
The bottom section refers to an hypothetical sample of more massive objects that should become available in the coming times.
We label this the ``future sample''.}
\begin{tabular}{|c|c|c|c|c|c}
\hline 
$z$& 0 & 0.5 & 1 &1.5 & 2 \\
\hline
Type & -  & LRGs & NVSS & QSOs & -\\
$b_{\rm{eff}}$& - & $1.9$ & $2$ & $2.3$ & -\\
$N_{\rm{gal}}$& - & $10\cdot 10^5$ & $1.4\cdot 10^5$ & $1.7\cdot 10^5$& - \\
\hline
Mass& - & 1.6-2.5 & 1-1.25 & 1-1.25& - \\
$b_1$ & -& 1.8  & 2  & 2.9 & -\\
$N_{\rm{halo}}$& - & $6\cdot 10^5$ & $4\cdot 10^5$ & $2.7\cdot 10^5$& - \\
\hline
 Mass &  6.25-9.5 & 4-6.25 & 4-6.25 & 4-6.25 & 2.5-4\\
$b_1$ & 1.8 & 2.5 & 4.0 & 4.9 & 5.9\\
$N_{\rm{halo}}$ & $1.7 \cdot 10^5$ & $2.1 \cdot 10^5$ & $1.1 \cdot 10^5$ & $0.4 \cdot 10^5$ & $0.4 \cdot 10^5$ \\
\hline
\end{tabular}
\label{tab:bins}
\end{table}
\begin{figure*}
\centering
\includegraphics[scale=0.35]{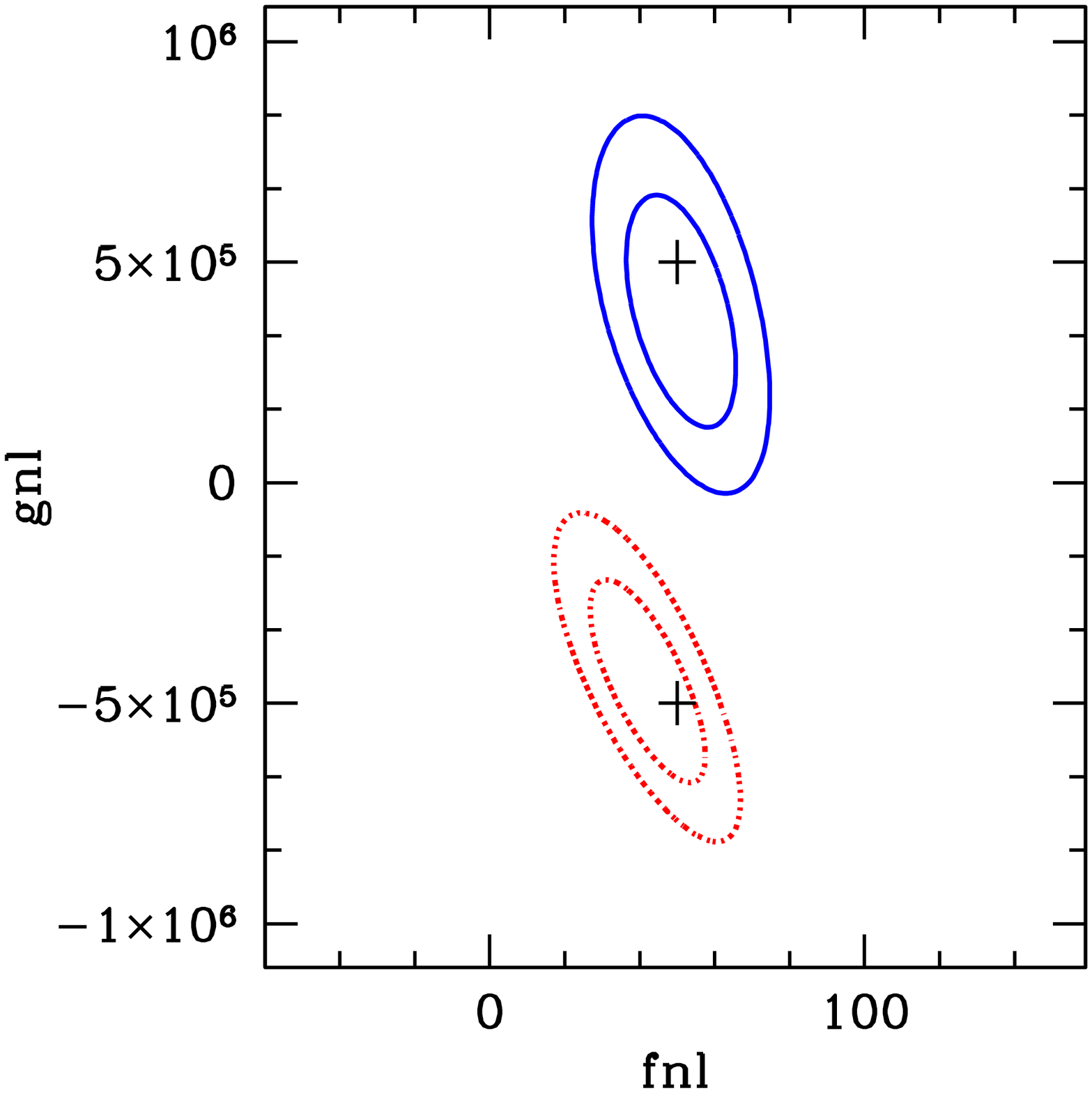}
\includegraphics[scale=0.35]{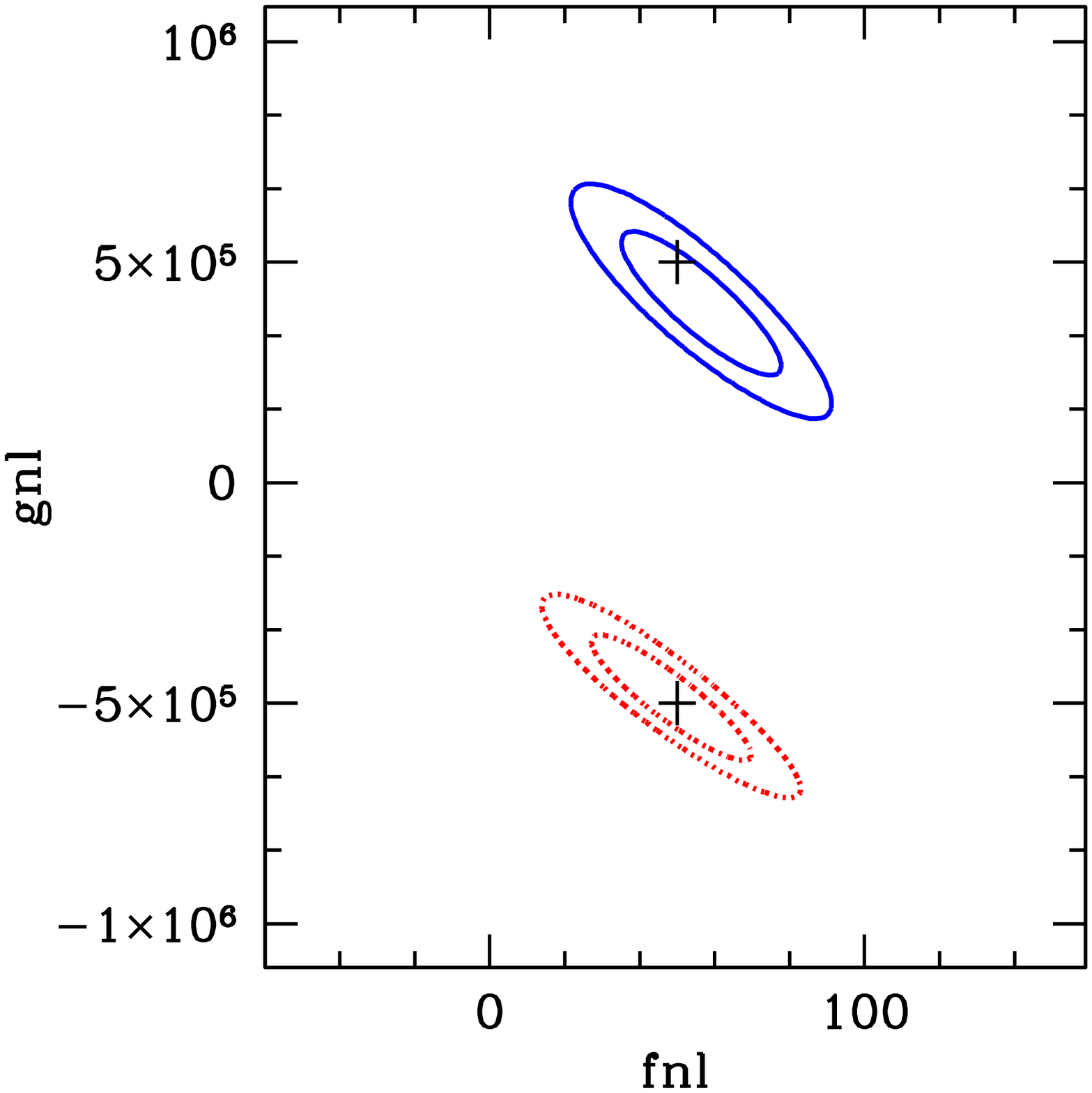}
\caption{Credibility intervals (68.3 and 95.4 per cent) for $\fnl$ and $\gnl$. Solid: positive $\gnl$, dotted: negative $\gnl$. The black crosses indicate the input values for $\fnl$ and $\gnl$. Left panel: using the current sample, right panel: using the future sample.}
\label{fig:cont_bmarg2}
\end{figure*}
\section{Parameter estimation}
\label{sec:est}
We employ the simulations to build mock observational data for $b(k)$ and use them to estimate the PNG parameters under different model assumptions. We consider two sets of halo mass and redshift bins as described in Table \ref{tab:bins}. 
The first set is chosen to closely resemble the observations used by \cite{2011JCAP...08..033X}. The second set covers a wider range in redshift and linear bias parameter $b_1$ and exemplifies datasets (for groups and clusters of galaxies)
that will be available in the near future (e.g. with the eROSITA mission plus spectroscopic follow up).
\subsection{Two-parameter model for PNG}
\label{ssec:fitb}
We first consider a model ($M_2$) with three free parameters, $\fnl$, $\gnl$ and $b_1$, and assume Gaussian errors on $b(k)$ to build their likelihood function. We adopt a flat prior within a finite region of parameter space (considerably more extended than the likelihood function) and marginalize the posterior distribution over $b_1$. Fig. \ref{fig:cont_bmarg2} shows the marginal posterior probability density for $\fnl$ and $\gnl$ by marking the 68.3 and 95.4 per cent credibility intervals for the model parameters. The two panels refer to the different sets of mass and redshift bins defined in Table \ref{tab:bins} (current data on the left, future data on the right). Different sets of curves in the same panel refer to the different fiducial models used to generate the initial conditions of the simulations. In all cases, the fiducial values for $\fnl$ and $\gnl$, indicated by the black crosses, lie within the 95.4 per cent credibility region region. Note that the current datasets are not expected to give tight constraints on the model parameters which are also degenerate. The credibility intervals shrink when more mass/redshift bins are used. The small rotation of the contour levels in the right panel is caused by including data from $z=2$, which significantly improves the constraints on $\gnl$, and only slightly weakens the constraints on $\fnl$ (with respect to the current dataset).
\subsection{One-parameter models for PNG}
\label{ssec:fits}
We now consider two simpler models by assuming that PNG is either purely quadratic ($\gnl \equiv 0$, $M_{1f}$) or purely cubic ($\fnl \equiv 0$, $M_{1g}$).
All the rest is left unchanged with respect to $M_2$, so that $M_2$, $M_{1f}$ and $M_{1g}$ form a set of nested models.

\begin{figure*}
\centering
\includegraphics[scale=0.34]{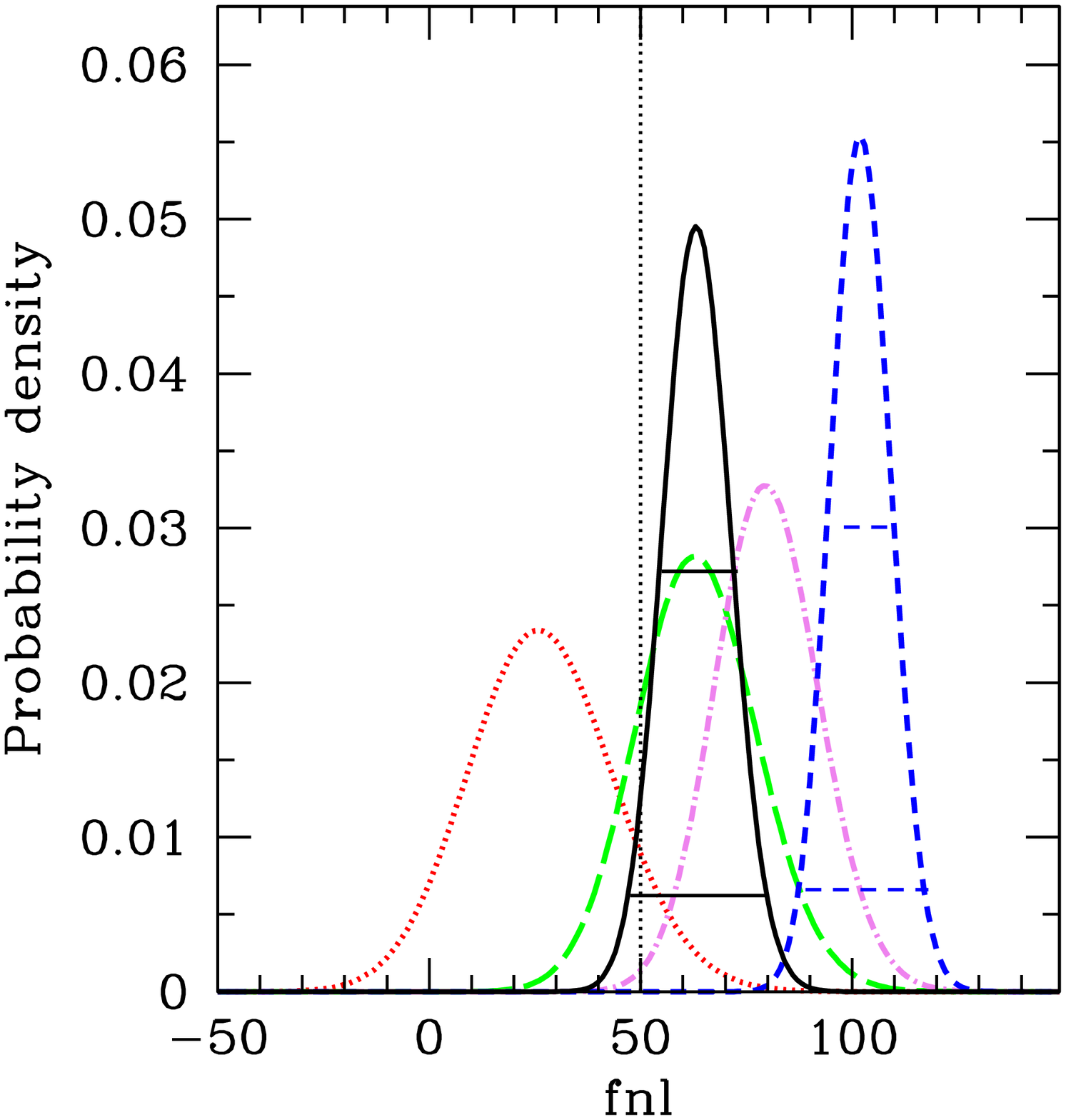}
\includegraphics[scale=0.34]{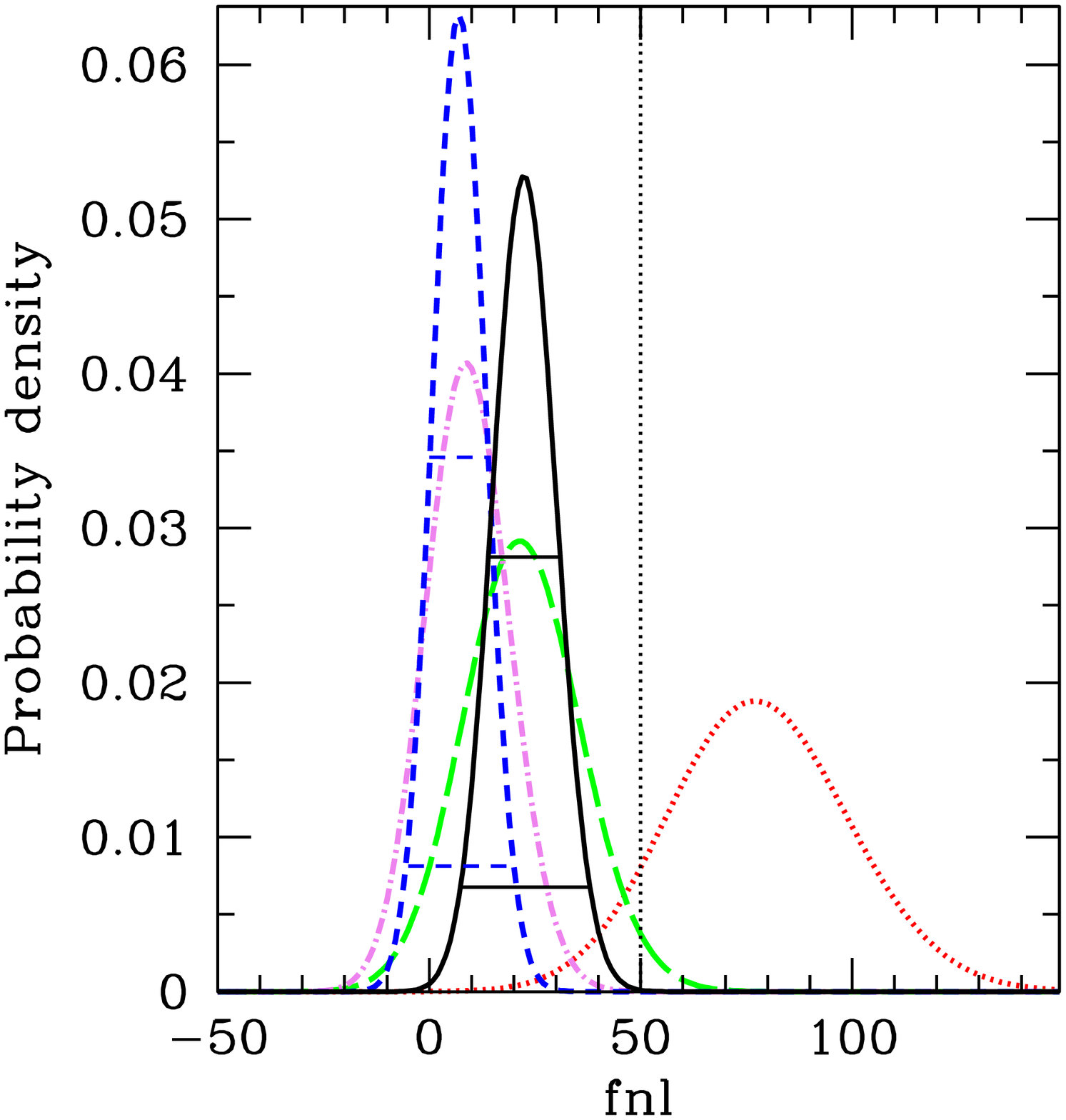}
\caption{Constraints on $\fnl$ for model $M_{1f}$ (where $\gnl \equiv 0$ is assumed a priori). Left: simulation with $\gnl=5\cdot 10^5$, right: simulation with $\gnl=-5\cdot 10^5$. Constraints from the 3 redshift bins in the current sample are shown with different line styles: $z=0.5$ dotted, $z=1$ long-dashed, $z=1.5$ dot-dashed. The posterior pdf obtained combining the 3 redshift bins is shown with a solid line. For comparison, we also show the combined posterior for the future dataset (short-dashed). The two horizontal lines indicate the 68.3 and 95.4 per cent credibility intervals for each of the combined posteriors. The vertical dotted line shows the input value $\fnl=50$.}
\label{fig:marg_g}
\end{figure*}

The marginal posterior probability distribution for $\fnl$ obtained combining the current dataset with model $M_{1f}$ is presented in Fig. \ref{fig:marg_g}. 
The left panel shows the result for the simulation with positive $\gnl$ and the right panel for negative $\gnl$. Different line styles indicate the individual redshifts bins: dotted, long-dashed, dot-dashed for $z=\{0.5,1,1.5 \}$, while the solid curve shows the combined probability for all redshifts. For the simulation with $\gnl=5\cdot 10^5$, the resulting 95.4 per cent credibility region (lower horizontal solid line) marginally includes the input value $\fnl=50$, indicated by the vertical dotted line, but the maximum is shifted towards higher values. When $\fnl$ and $\gnl$ have opposite sign, the maximum a posteriori value is biased low, and the input value is excluded at high significance. In both cases, there is an apparent redshift dependence of $\fnl$, and this trend depends on the sign of $\gnl$. Thus, if a redshift dependence was ever detected from observations, it may just indicate that the wrong model for PNG is assumed. We note that the results in \cite{2011JCAP...08..033X} actually hint at a redshift dependence of $\fnl$ although the error bars are quite large. If taken at face value, the ordering of the estimates would suggest that $\gnl<0$.

Additionally, Fig. \ref{fig:marg_g} includes the combined probability for the 5 mass and redshift bins of the future dataset for comparison (short-dashed). The shift away from the input value is qualitatively the same, but \textit{larger} for the future dataset. This can be understood by comparing the degeneracy between $\fnl$ and $\gnl$ in Fig. \ref{fig:cont_bmarg2}.

\begin{figure*}
\centering
\includegraphics[scale=0.34]{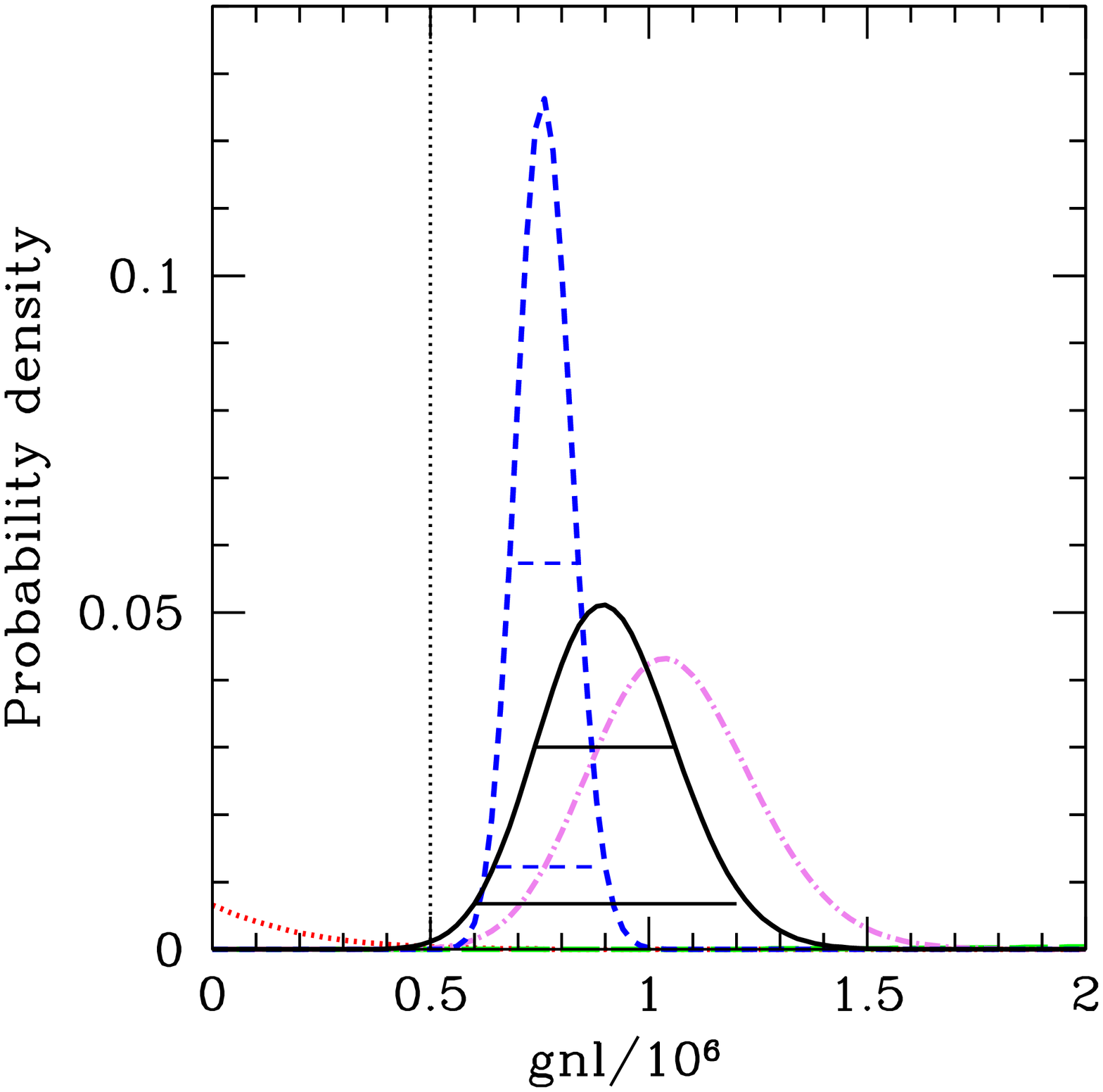}
\includegraphics[scale=0.34]{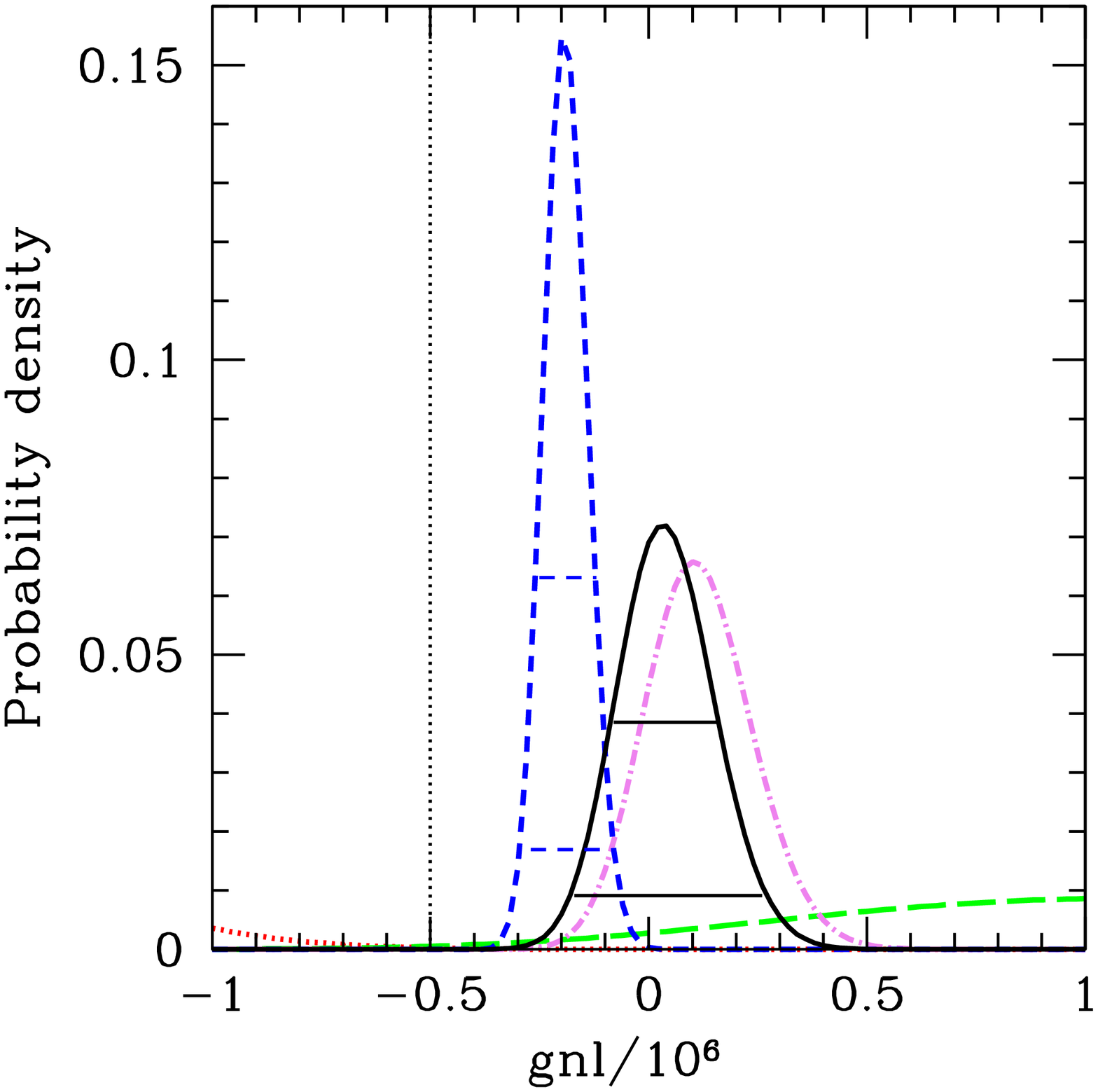}
\caption{Constraints on $\gnl$ for model $M_{1g}$ (where $\fnl \equiv 0$ is assumed a
priori). Left panel: simulation with $\gnl=5\cdot 10^5$, right panel: simulation with $\gnl=-5\cdot 10^5$. Line styles are the same as in Fig. 2. The vertical dotted line shows the input value $\gnl=\pm \, 5 \cdot 10^5$. Note that the long-dashed curve ($z=1$) seems to be missing in the left panel because the corresponding probability density is too low.}
\label{fig:marg_f}
\end{figure*}

The marginal posterior probability distribution for $\gnl$ under model $M_{1g}$ is shown in Fig. \ref{fig:marg_f}. As in the previous Figure, the two panels show the results for the simulations with $\gnl=\pm \, 5\cdot 10^5$. In both cases, the maximum a posteriori estimate is biased high, and the input value lies well outside the 95.4 per cent credibility regions. The estimates from the different redshifts are almost incompatible with each other, and there is no apparent trend with $z$ as seen in model $M_{1f}$. For the case of negative $\gnl$, the estimation procedure favours a Gaussian model although both input PNG parameters are substantially different from 0. The combined probability of the future dataset is shown by the short-dashed curve. Here, the shift away from the input value for $\gnl$ is \textit{smaller}, and the constraints are tighter than for the current dataset, in accordance with the rotation of the contours in Fig. \ref{fig:cont_bmarg2}.
\subsection{Model selection}
\label{ssec:model}
Our results show that the choice of the underlying model for PNG has a significant effect on the estimates of $\fnl$ and $\gnl$. Only model $M_2$ returned estimates which were in agreement with the fiducial values, while setting a parameter
to zero a priori (as often done in the literature) produced biased results. Here we investigate if the data themselves can be used to establish which parameterisation should be preferred. 

Given a set of data $D$, 
the relative evidence in favour of a model with respect to a second one
can be quantified in terms of the Bayes factor (e.g. \citealt{Kass1993}, \citealt{2008ConPh..49...71T})
\be
B_{ab}=\frac{P(D|M_a)}{P(D|M_b)}=
\frac{\int \mathcal{L}_a(D|M_a, \theta_a) \, \pi_a(\theta_a)\, \rm{d}\theta_a}{\int \mathcal{L}_b(D|M_b,\theta_b)\, \pi_b(\theta_b) \, \rm{d}\theta_b} .
\label{BErat}
\ee
Here $\mathcal{L}_i(D|M_i, \theta_i)$ gives the probability of getting the data $D$ given the model $M_i$ 
with parameters $\theta_i$ (i.e. $\mathcal{L}_i$ is proportional to the likelihood function for the model 
parameters), $\pi_i(\theta_i)$ is the prior probability density for the model parameters and $P(D|M_i)$
denotes the Bayesian evidence for model $M_i$ (i.e. the probability of getting the data under model $M_i$
after marginalizing over the values of the model parameters).

We find that $M_{2}$ is strongly favoured over $M_{1f}$ and $M_{1g}$, with a Bayes factor 
$> 30$, except for one case\footnote{Using the current sample, $M_{1f}$ is only disfavoured by a factor of 9 (3) for the simulation with positive (negative) $\gnl$.}. Based on the Jeffreys scale \citep{jef}, this provides ``very strong'' evidence in favour of $M_2$. We thus recommend the use of Bayes factors in future determinations of PNG from observational data based on galaxy 2-point statistics.  
\section{Conclusion}
\label{sec:conc}
Using two sets of $N$-body simulations starting from non-Gaussian initial conditions of the local type
with $\{\fnl, \gnl\}=\{50,\, \pm\, 5 \cdot 10^5\}$, we have investigated the bias of dark-matter halos on large scales. Motivated by the widespread practice of considering only one-parameter models for PNG to fit observational data, we have studied the effect of model assumptions on the determination of model parameters based on 2-point statistics of the galaxy distribution. We have considered two sets of mass and redshift bins: the first one is chosen to emulate the current observational samples, while the second approximates data from future cluster surveys. Our main conclusions can be summarized as follows:
\begin{itemize}
\item Fitting the mock data with the input (two-parameter) PNG model gives unbiased results although
estimates for $\fnl$ and $\gnl$ turn out to be degenerate.
This degeneracy can be at least partially broken using higher-order statistics of galaxy clustering, like the bispectrum. 
\item If we assume a purely quadratic model ($\gnl=0$), the maximum a posteriori estimate for $\fnl$ is biased high (low) for the simulations with $\gnl=5 \cdot 10^5$ ($\gnl=-5 \cdot 10^5$), respectively. The input value of $\fnl=50$ is almost always excluded at very high confidence. If we consider each redshift bin separately, a spurious systematic shift of the $\fnl$-estimate with $z$ appears, which also depends on the sign of $\gnl$. If seen in observations, this effect could actually indicate the necessity for a more complex model.
\item If we assume a purely cubic model ($\fnl=0$), the maximum a posteriori estimates for $\gnl$ are artificially shifted to larger values, which differ from the input value much more than the broadness of the posterior distribution. When the fiducial PNG parameters have opposite sign, the two contributions to the scale-dependent bias can cancel out, leading to a false null detection.
\item
Our analysis could be extended by considering even higher-order terms in the expansion of equation (\ref{png}). However, this phenomenological approach adds degrees of freedom which are not necessarily independent because they may be related to each other through the underlying physics. Alternatively, one could test specific inflationary models and put direct constraints on their properties (e.g. coupling constants) that determine the PNG parameters.
\item 
Given a dataset, statistical model selection techniques are capable to identify the optimal
PNG parameterization and thus avoid estimation biases due to incorrect simplifying assumptions.
The Bayes factor very strongly favours the two-parameter model over a purely quadratic or purely cubic model for our simulations. 
It is thus important to apply this model selection to future constraints on PNG from observational data.
\end{itemize}

\section{Acknowledgements}
\label{sec:ac}
NR would like to thank Annalisa Pillepich for help with \textsc{Gadget-2}. The simulations were performed at the Leibniz-Rechenzentrum (LRZ) in Garching, Germany. We acknowledge support through the SFB-Transregio 33 ``The Dark Universe" by the Deutsche Forschungsgemeinschaft (DFG).
\vspace{-1cm}
\bibliography{db}
\bibliographystyle{mn2e}

\end{document}